# Transmission-and-Distribution Frequency Dynamic Co-Simulation Framework for Distributed Energy Resources' Frequency Response

Wenbo Wang, *Member, IEEE*; Xin Fang, *Senior Member, IEEE*; Hantao Cui, *Senior Member, IEEE*; Fangxing Li, *Fellow, IEEE*

*Abstract*— The rapid deployment of distributed energy resources (DERs) in distribution networks has brought challenges to balance the system and stabilize frequency. DERs have the ability to provide frequency regulation; however, existing dynamic frequency simulation tools—which were developed mainly for the transmission system—lack the capability to simulate distribution network dynamics with high penetrations of DERs. Although electromagnetic transient (EMT) simulation tools can simulate distribution network dynamics, the computation efficiency limits their use for large-scale transmission-and-distribution (T&D) simulations. This paper presents an efficient T&D dynamic frequency co-simulation framework for DER frequency response based on the HELICS platform and existing off-the-shelf simulators. The challenge of synchronizing frequency between the transmission network and DERs hosted in the distribution network is approached by detailed modeling of DERs in frequency dynamic models while DERs' phasor models are also preserved in the distribution networks. Thereby, local voltage constraints can be respected when dispatching the DER power for frequency response. The DER frequency responses (primary and secondary)—are simulated in case studies to validate the proposed framework. Lastly, fault-induced delayed voltage recovery (FIDVR) event of a large system is presented to demonstrate the efficiency and effectiveness of the overall framework.

## I. INTRODUCTION

DISTRIBUTED energy resources (DERs) are being rapidly deployed in distribution networks, which brings new challenges to balance the power system and stabilize the system frequency [1]. The DER power outputs, if not optimally managed, can impact not only the local distribution voltage profile but also the transmission system power balance and can increase frequency fluctuations [2], [3]. Frequency regulation services—including primary frequency response (PFR), secondary frequency response (SFR), and tertiary frequency response—are used to maintain real-time system balance and frequency stability. DERs, equipped with advanced control strategies, have the capability to provide frequency regulation services [4], [5]. The recent Federal Energy Regulatory Commission Order 2222 [6] stipulates that electricity markets should remove all market access barriers to DERs to allow them to participate in the energy, capacity and ancillary services markets; therefore, to better understand and use DER frequency regulation capabilities, the dynamic frequency response of DERs should be modeled in the system dynamic simulation.

On the one hand, existing dynamic simulation tools, such as GE PSLF and PTI PSS/E [7], [8], however, were developed mainly for transmission frequency dynamic analysis where positive sequence model (balanced three-phase) are assumed. This assumption is largely valid for the transmission network. On the other hand, the distribution networks hosting DERs is normally three-phase unbalanced, and the DERs' power outputs from the frequency response should not violate the local constraints. Therefore, how to integrate the distribution network DER dynamics into the system frequency dynamic simulation attracts increasing research attention.

Electromagnetic transient (EMT) simulation tools can simulate both transmission and distribution (T&D) network dynamics; however, the full EMT simulation for T&D networks requires extensive simulation time—even for a medium-size network [9]. Using the full EMT simulation to simulate large-scale T&D networks is unrealistic. In [10], a hybrid EMT and phasor-domain simulation model was proposed to accelerate the EMT simulation for T&D networks. The EMT simulation was accelerated by switching between the detailed EMT simulation and the phasor-domain simulation. In [11], an integrated T&D system power flow and dynamic simulation was proposed. Both T&D networks were modeled in detail with three sequences and three phases. These models were accurate for the EMT simulation; however, the simulation times were still extensive for the large-scale T&D network simulation. In addition, the characteristic of the transmission system three-phase balance was not leveraged, and the existing high-performance transmission frequency dynamic simulation tools were not incorporated. A three-phase unbalanced transient dynamic simulation model was proposed in [12] for the distribution network with synchronous generators. This model used the electromechanics transient model in the distribution network and microgrids. Also, the transmission frequency dynamics were not discussed. In [13], T&D dynamic co-simulation models with parallel computation and series computation between different subsystems were proposed. The computational performance of this co-simulation framework was not discussed. Reference [14] proposed a hybrid simulation tool to study the impacts of distributed photovoltaics (PV) in distribution networks. Distributed PV was modeled with the

¹This work was authored by Alliance for Sustainable Energy, LLC, the manager and operator of the National Renewable Energy Laboratory for the U.S. Department of Energy (DOE) under Contract No. DE-AC36-08GO28308. Funding provided by the U. S. Department of Energy Office of Electricity Advanced Grid Research and Development program. The U.S. Government retains and the publisher, by accepting the article for publication, acknowledges that the U.S. Government retains a nonexclusive, paid-up, irrevocable, worldwide license to publish or reproduce the published form of this work, or allow others to do so, for U.S. Government purposes. The views expressed in the article do not necessarily represent the views of the DOE or the U.S. Government.

Wenbo Wang and Xin Fang are with the National Renewable Energy Laboratory, Golden, CO, 80401, USA.

Hantao Cui and Fangxing Li are with the Department of EECS, University of Tennessee, Knoxville, TN, 37996, USA.

EMT model to study its fast dynamics, and the distribution network was modeled with a quasi-static time-series (QSTS) simulation in OpenDSS [15]. Although this model considered PV dynamics with the EMT simulation in the distribution network, how PV dynamics impact the transmission frequency response was not studied.

DERs provides frequency regulation by adjusting their active power output, meanwhile DERs may be required by local operators to adjust reactive power to maintain certain voltage ranges. This provides opportunities for DERs to maximize their utilizations. Therefore, as a first step, to account for DER dynamic frequency regulation response with high computational efficiency, a T&D frequency dynamic co-simulation framework is proposed in this paper. It leverages the high performance transmission dynamic simulation tool (ANDES) [16] and the distribution network solver (OpenDSS) [15]. The co-simulation platform will be built on the Hierarchical Engine for Large-scale Infrastructure Co-Simulation (HELICS) [17], [18] to establish the efficient co-simulation flow between the transmission dynamic simulation and the distribution QSTS power flow simulation. To synchronize the frequency between the transmission network and the DERs hosted in the distribution network, DERs are modeled in the transmission frequency dynamic simulation with the detailed frequency dynamic model. The DER power outputs are then sent to the distribution network's power flow simulation through HELICS. Consequently, the DER frequency responses are considered in both the transmission frequency dynamic simulation and the distribution network power flow simulation. Because both the transmission dynamic simulation and the distribution QSTS simulation are very fast, the proposed T&D dynamic co-simulation framework is computationally efficient. The main contributions of this paper are:

1) A general-purpose T&D frequency dynamic co-simulation framework is developed to study the DER dynamic frequency responses in large-scale T&D networks.
2) A novel implementation to synchronize DER PFR and SFR and the transmission system frequency by modeling DER dynamics in the transmission model while the DER QSTS model are also preserved in the distribution system.
3) The DER power intermittency and local voltage constraints are considered in the PFR and SFR provision. The real time maximum available power of the DERs is considered by an optimization formulation.
4) The T&D frequency dynamic co-simulation framework is further demonstrated on a detailed generation outage simulation of a transmission network test case with 10 distribution networks, including a large 8,500-node distribution system. The fault-induced delayed voltage recovery (FIDVR) case is studied as well.

This paper is organized as follows: Section II presents the overall T&D frequency dynamic co-simulation framework in HELICS. Section III introduces the T&D network frequency dynamic model with DERs. Section IV performs the case study to demonstrate the PFR and SFR of the DERs. The computational performance of the proposed T&D frequency dynamic co-simulation framework is shown as well. Section V concludes the paper.

## II. TRANSMISSION-AND-DISTRIBUTION FREQUENCY DYNAMIC CO-SIMULATION FRAMEWORK

The dynamic T&D co-simulation framework developed for the DER dynamic frequency response is based on the HELICS platform and off-the-shelf power system simulation software. This section introduces the components of the framework and develops the interfacing requirements.

### A. Brief Description of HELICS

HELICS is an open-source, cyber-physical, co-simulation framework for energy systems. It is designed to integrate simulators of transmission, distribution, and communications domains to simulate regional and interconnection-scale power system behaviors. A few key concepts of HELICS that are relevant here are introduced below; for more details, see [17], [18]:

- Brokers maintain synchronization in the federation (i.e., many federates) and facilitate message exchange among federates.
- Federates are running simulation instances of individual systems, sending and receiving physical and control values to and from other federates.
- Simulators are executable—that is, they can perform some analysis functions. In this context, for example, they are the transmission simulator ANDES and distribution simulator OpenDSS. Note that the terms *federate* and *simulator* are used interchangeably in this paper.
- Messages are the information passed between federates during the execution of the co-simulation. The messages exchange is realized through either defining subscriptions and publications or federate-to-federate endpoint communications.

### B. T&D Networks with DERs and Interface

The T&D simulators can be run in different federates (for example, different Python files), the synchronization is maintained by a HELICS broker, and the information exchange needs to be defined next.

The transmission system simulator performs the time domain simulation (frequency, i.e., electromechanical dynamics), whereas the distribution system simulator performs the QSTS simulation. Fundamentally, these simulators perform calculations of the differential-algebraic equations (DAE) for the transmission system, the algebraic equations for distribution systems, the algebraic equations for thermal controllers, and the algebraic equations for DER aggregators. The co-simulation data exchange among the simulators in terms of simulation time is configured as loosely coupled (i.e., with one iteration between simulators) so that the co-simulation is robust, and the impacts can be found in [13]. For demonstration, only one transmission and one distribution network are shown in Fig. 1. Note that the HELICS platform can simulate multiple distribution networks connected to the transmission network in parallel. The arrow pointing to the right denotes the simulation time; the dashed rectangles denote the changing states (in terms of simulation time) of the T&D simulators. This is also true when both the thermal controller and DER controller simulators are added.

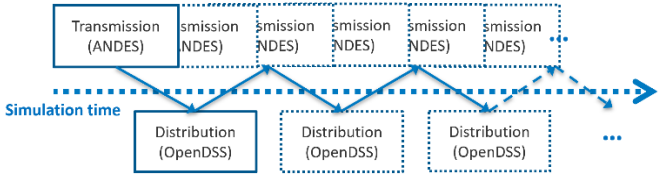

Fig. 1. Simulation flow demonstration in terms of simulation time.

The detailed information exchanged through each simulator is shown in Fig. 2, which can almost be seen as a snapshot of Fig. 1. The arrows here denote the information exchange directions, with the exchange step time displayed as well. Blue represents the physical power system simulators and values, whereas orange represents the communications simulators and signals. The transmission simulator and the distribution simulator exchange the physical information every second. The transmission internal simulation time step is two cycles (33.3 milliseconds). The transmission dynamic simulator sends the system frequency and area control error (ACE) signals to the thermal controller and the DER aggregator every 0.5 second. The controllers send their automatic generation control (AGC) control signal to the transmission dynamic simulator every 4 seconds.

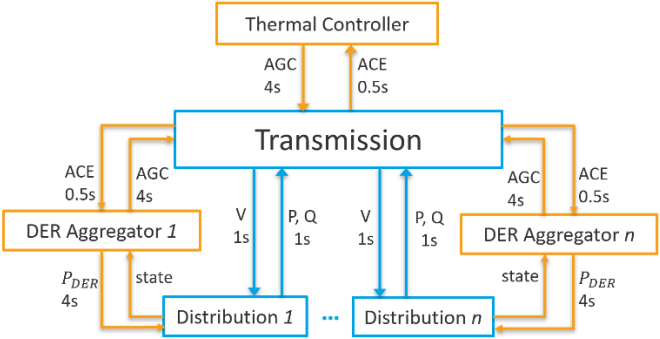

Fig. 2. Information exchange and frequency.

More specifically, because the transmission simulator is in a per-unit positive-sequence equivalent calculation, the voltage results are then sent as a three-phase balanced voltage source to the distribution networks. For the distribution system simulators, the system three-phase power injection/withdraw (at the substation) can be transformed into the positive-sequence power injection using the formulation (1)–(3) [19] and then send back to the transmission simulator:

$$S_i^+ = \boldsymbol{T} \boldsymbol{S}_i^{abc} \quad (1)$$
$$S_i^+ = P_i^+ + jQ_i^+ \quad (2)$$
$$\boldsymbol{T} = [1/3 \quad 1/3 \quad 1/3] \quad (3)$$

where $S_i^+$ is the power at bus $i$ in the transmission positive-sequence dynamic model; and $\boldsymbol{S}_i^{abc}$ is the three-phase power of the distribution network connecting the transmission bus $i$.

### C. Simultaneously Handling DER Dynamic and QSTS Models

Because the distribution network simulators exchange the physical information with the transmission dynamic simulator every second, the fast frequency dynamics—such as the synthetic inertia response and PFR—cannot be accurately captured with this large time step of the information exchange; therefore, the DER dynamic model should be included in the transmission simulator so that the DER dynamic frequency response—especially the fast synthetic inertia and PFR—can be included in the transmission dynamic simulation. Most DERs are hosted in distribution systems, however, where local voltage needs to be maintained in the range from 0.95–1.05, along with distribution line rating limits. To account for these local requirements, the DER static power flow models are also considered in the distribution simulators. This treatment is then completed by adjusting the overall power injection at the substation from the distribution simulators, as in (4):

$$\boldsymbol{S}_i^{abc} = \boldsymbol{S}_{i,with\ DER}^{abc} + \sum \boldsymbol{S}_{i,DER}^{abc} \quad (4)$$

where $\boldsymbol{S}_{i,with\ DER}^{abc}$ is the distribution abc three-phase net load, including the DER power outputs (the distribution is connected to transmission bus $i$); and $\boldsymbol{S}_{i,DER}^{abc}$ is the DER power outputs in the abc three phases.

### D. Co-Simulation Integration with HELICS

The HECLIS simulation platform can accommodate the aforementioned simulators; then the synchronization of the simulation time among different simulators is controlled implicitly by a broker, and the information exchanges are realized by either subscriptions/publications or end point communications [18]. The schematic structure is shown in Fig. 3. The HELICS command line interface (cli tool) can be used in a Terminal script to launch the co-simulation (e.g., running all the simulators simultaneously) [18]. Note that the communications variation regarding both the latency and packages dropping can be modeled in the end point; therefore, the cyber-physical interaction can be simulated in this platform. The co-simulation platform includes HELICS, ANDES, and OpenDSS. All are open-source packages; therefore, the proposed dynamic T&D co-simulation platform can be used without any commercial license limitations.

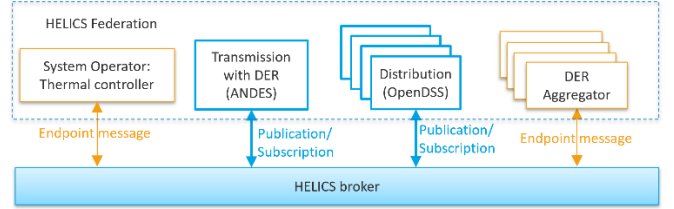

Fig. 3. Co-simulation framework structure in HELICS.

## III. MODELING OF T&D NETWORK FREQUENCY DYNAMICS WITH DERs

This section describes the T&D network dynamics with DERs hosted in the distribution networks while respecting local voltage constraints. The dynamic model of distributed PV is shown in Fig. 4. More details about this model can be found in [20]. A limit for PV's available power, $P_{mppt}$, based on maximum power point tracking (MPPT), is added to limit PV's real-time total power output.

In Fig. 4, $P_{ref}$ is the reference power, updated every 5 minutes, which is obtained from the system operator's real-time economic dispatch. Its value is kept constant in the 5-minute interval. $P_{drp}$ is the PFR power set point from the droop response. $P_{ext}$ is the SFR power set point, which is obtained from the system AGC control signal every 4 seconds. The total active power output of $P_{ref}$, $P_{drp}$, and $P_{ext}$ should not exceed PV's maximum available power, $P_{mppt}$. This will be introduced with more details in the following subsections.

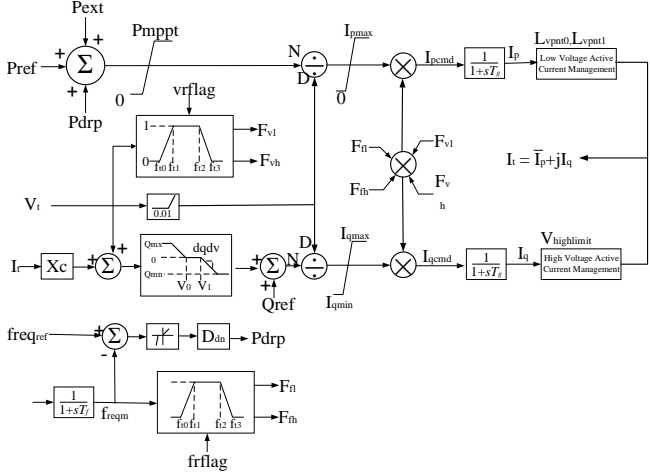

Fig. 4. Distributed PV power plants generic model.

### A. DER Frequency Response Modeling

#### 1) Droop Control for PFR

The dynamic model of distributed PV in Fig. 4 includes droop control for PFR. When the frequency drop is larger than the PFR deadband, PV will change its active power output accordingly. An additional power output, $P_{drp}$, will be included for its PFR.

$$P_{drp} = \begin{cases} \frac{(60-db_{UF})-f}{60} D_{dn}, f < 60 \\ \frac{f-(60+db_{OF})}{60} D_{dn}, f > 60 \end{cases} \quad (5)$$

where $db_{UF}$ and $db_{OF}$ are the underfrequency and overfrequency deadband; $D_{dn}$ is the per-unit power output change to 1 per-unit frequency change (frequency droop gain).

#### 2) SFR through AGC

The SFR is enabled by an AGC model, which is implemented through two components: an area-level model that calculates the ACE from (*) representing the system active power imbalance; and a plant-level control model that receives the ACE signal and sets the reference power for each plant.

Assuming for simplicity that there is only one area in the simulation and no interchange with other areas, the area-level AGC model is implemented as shown in Fig. 5. The ACE is defined according to the North American Electric Reliability Corporation technical document [21] with the interchange metering error ignored, i.e.:

$$ACE_{tt} = -10B(f_{reqm,tt} - f_0) \quad (6)$$

where $tt$ is the AGC time interval index; $ACE_{tt}$ is the ACE at the AGC interval $tt$; $f_{reqm,tt}$ is the measured system frequency at the AGC interval $tt$; $f_0$ is the system reference frequency (60 Hz); and $B$ is the frequency bias in MW/0.1 Hz. Eq. (*) represents the area's active power imbalance estimated from the system frequency deviation. In the implementation, a frequency error tolerance deadband, $f_{db}$, is introduced to eliminate the unnecessary movement of the generation set points. A proportional integral (PI) control is applied to the ACE signal to generate the actual area control signal that will be passed on to individual generators. $K_p$ and $K_i$ are the coefficients of the PI controller. The ACE signal and the area control signal are updated every 4 seconds to represent their discrete nature in the field. The ACE signal is then passed on to each AGC generator considering the unit's participation factor such that the individual AGC power plant's power reference is updated accordingly. The participation factor of each unit's AGC response is decided by the real-time economic dispatch. In Fig. 5, $\beta_i$ is the i-th unit's participation factor.

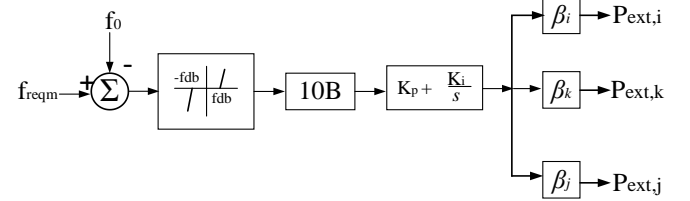

Fig. 5. AGC model.

### B. Transmission Frequency Dynamic Simulation with DERs

The transmission system frequency dynamic simulation is performed with ANDES, an open-source Python-based dynamic simulation library [22]. ANDES used a hybrid symbolic-numeric framework for the system dynamic modeling and simulation. The system dynamics can be modeled as a set of mass-matrix DAEs [23]:

$$\boldsymbol{M\dot{x}} = \boldsymbol{f}(\boldsymbol{x}, \boldsymbol{y}, \boldsymbol{u}) \quad (7)$$
$$\boldsymbol{0} = \boldsymbol{g}(\boldsymbol{x}, \boldsymbol{y}, \boldsymbol{u}) \quad (8)$$

where $\boldsymbol{f}$, $\boldsymbol{g}$ are the differential and algebraic equations, respectively. $\boldsymbol{x}$, $\boldsymbol{y}$, and $\boldsymbol{u}$ are the state, algebraic variables, and inputs; and $\boldsymbol{M}$ is the mass matrix.

### C. Distribution QSTS Power Flow Simulation with DER SFR Headroom Estimation

The distribution system QSTS power flow simulation is performed with OpenDSS. To account for the local voltage constraint that might be incurred by DER from frequency response, the DERs are modeled in distribution systems as well but with a phasor model. As discussed in Subsection II-C, this will ensure that the DERs respect the local constraints and fit into the overall co-simulation framework.

For DERs to provide frequency response, for a certain time step, distribution system operators or DERs aggregators submit the DER headroom to the transmission system operators. This headroom is estimated through a fast (linear) optimization scheme, as in (9)–(12). The objective function (9) maximizes the total output of DERs in a specific distribution system while respecting constraints, including voltage and thermal limits (11), (12), as well as an equality constraint (10) of the voltage-power sensitivity matrix (VSM, denoted by $J_{VSM}$), which can be seen as power flow equations linearized at certain system states. VSM is obtained based on [24], but it focuses only on the DER nodes and active power: at each time step, perturb power injections at the nodes that are connected with DERs, one at a time, until exhausting all the DER nodes.

$$\max(\boldsymbol{1}^T P_{DERs}) \quad (9)$$
$$\text{s.t.} \quad J_{VSM} \Delta P_{DERs} = \Delta V \quad (10)$$
$$\underline{V} < V_{base} + \Delta V < \overline{V} \quad (11)$$
$$\underline{I} < I < \overline{I} \quad (12)$$

where $\boldsymbol{1}$ is a column vector with all elements being 1; $P_{DERs}$ is a column vector with size $m\times 1$ that contains the $m$ DER outputs; $\Delta P_{DERs}$ represents the change in DER outputs; $\Delta V$ represents the change in voltage at all nodes (assume $n$ nodes) in a feeder; $J_{VSM}$ denotes the sensitivity matrix with size $n\times m$; $V_{base}$ is the voltage value from the current time step; and $I$ represents the current flow in the circuits.

The output from this optimization scheme is the total maximal headroom calculated by (13), then it is sent to the transmission system operators:

$$P_{headroom} = max(\sum P_{DERs}) - \sum P_{DERs}^{Base} \quad (13)$$

where $P_{DERs}^{Base}$ is the DER output at the current time step. The transmission system operator then considers these limits, which is shown in (14):

$$P_{DER} = min\ (P_{DER}^{PFR} + P_{DER}^{SFR} + P_{DER}^{Base}, P_{mppt}, P_{caps}, P_{DER}^{Base} + P_{headroom}) \quad (14)$$

where $P_{caps}$ denotes the capacity ratings for DERs.

## IV. CASE STUDIES

This section further corroborates the proposed T&D dynamic co-simulation framework through case studies. For simplicity and to illustrate the effectiveness of the framework, the first test system is assembled based on the IEEE 14-bus transmission system [16] with 2 load buses connected with detailed feeder models, including the IEEE 34-bus distribution network and the IEEE 8,500-node distribution network [25]. This demonstrates the superiority of the framework in computational performance (parallel computing among distribution federates).

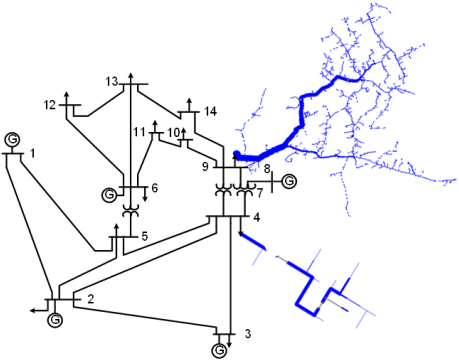

Fig. 6. Integrated T&D network with IEEE 14-bus system.

### A. T&D Network with IEEE 14-Bus Tranmission System

The integrated T&D network consists of the IEEE 14-bus system and 2 distribution feeders (8,500-node and 34-bus), as shown in Fig. 6, with Bus 4 connected with the IEEE 34-bus feeder and Bus 9 connected with IEEE 8,500-node feeder. Each feeder hosts 10 DERs with nameplate ratings of 900 kVA. Each feeder has an aggregator coordinating the DER AGC response in its own distribution feeder. The DER current dispatched active power outputs are 500 kw. The total time of the domain dynamic simulation is 60 s. The transmission dynamic parameters can be found in [16]. Gen 5 in the original IEEE 14-bus system reduces its power output from 35 MW to 25 MW. The total active power output of the DERs is 10 MW, with 0.5 MW for each DER. Here, for model simplicity, all DERs are assumed to be distributed PV; other types of DERs can be modeled as well.

### B. DER AGC Response with Load Variation

This subsection studies the DER's SFR AGC response in the proposed T&D dynamic co-simulation model. In normal operation, it is assumed that loads vary randomly with 2% standard deviation in the simulated time; this variation is shown as the load multiplier in Fig. 7. The total simulation time is 60 seconds. DERs can provide SFR based on the AGC signals provided by the DER aggregators. The system sends the aggregated AGC control signal to each aggregator. Then the aggregators disaggregate the AGC signal to individual DERs. In this study, the AGC signal is sent to the individual generator or DER every 4 seconds. Because the participation factor of the AGC response is decided in the economic dispatch (every 5 minutes), which is not included in this T&D dynamic co-simulation model, it is assumed that 20 DERs provide 10% of the AGC response, with each DER providing 0.5% of the AGC response. The rest of the thermal units provide 90% of the AGC response.

TABLE I. STATISTICS OF FREQUENCY AND ACE

| Item | Frequency (Hz) | ACE (MW) |
|---|---|---|
| Mean | 59.999 | -0.172 |
| Std | 0.026 | 4.174 |
| Min | 59.936 | -10.371 |
| Max | 60.054 | 8.833 |

Table I summarizes the statistics metrics of the frequency and ACE. The mean frequency is close to 60 Hz. The maximum and minimum frequency deviation is within 0.065 Hz. This means that the system frequency performance is good under the load variation. Fig. 8 shows the system frequency and the probabilistic distribution of the frequency across the simulated time. The frequency varies within a small range around +/- 0.05 Hz.

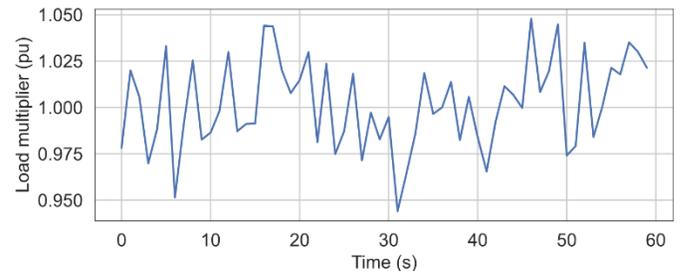

Fig. 7. Variation of load multiplier in the simulated time.

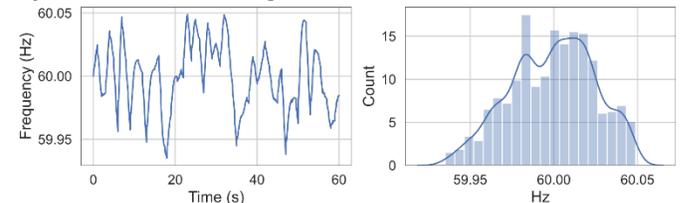

Fig. 8. System frequency and distribution.

Fig. 9 shows the DER AGC signal provided by DER aggressors (example from one DER). When the DERs do not provide SFR services, their AGC signal will be zero. When the

DERs provide SFR, their AGC signal will change based on the system ACE. This figure also demonstrates that the DER AGC signal changes every 4 seconds. Fig. 10 further demonstrates that the output of the DER varies according to its AGC signals, and this output considers local voltage constraints that are based on the optimization scheme in Subsection III-C and DER maximum available power (MPPT, DER capacity ratings). In this study, the DER $P_{mppt}$ is a 1-second time-series input data. The maximum power from the VSM is calculated every 10 seconds. Fig. 10 shows that the DER output—including its SFR response at all times—is less than $P_{mppt}$ (decided by irradiation) and the maximal value limited by local voltage constraints (see VSM max); therefore, both the available power variation resulting from solar intermittency and the local distribution voltage limits can be respected when DERs provide the SFR to the transmission system.

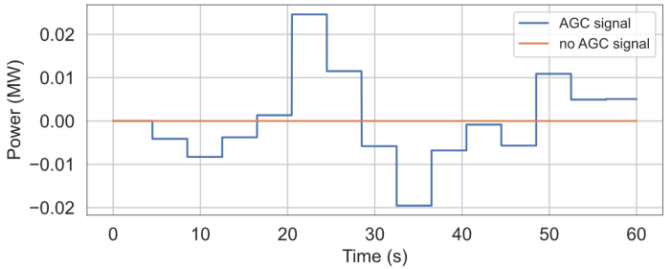
Fig. 9. Example of DER AGC signal from a DER aggregator.

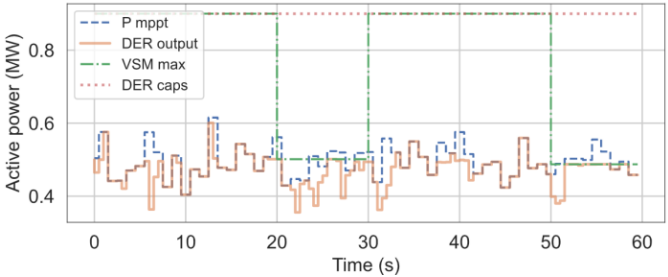
Fig. 10. Example of DER AGC signal from a DER aggregator.

Fig. 11a shows the overall voltage for the feeder 34-bus (connected with Bus 4 of the 14-bus transmission network), and Fig. 11(b) shows the 8500-node (connected with Bus 9 of the transmission network). The solid line shows the average voltage within the feeder for the simulated time, and the shaded area denotes one standard deviation from the average. The dashed lines mark the minimum and maximum of the feeder voltages. It can be observed that two feeder voltages are within 0.95~1.05 p.u. when the system load varies.

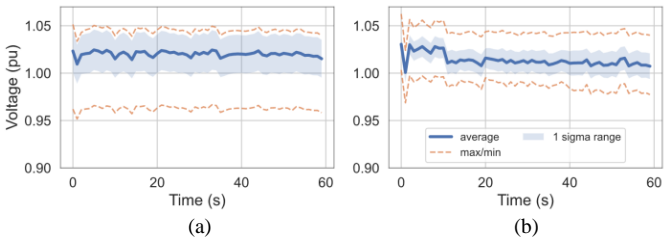
Fig. 11. Voltage plots of (a) feeder 34-node; (b) feeder 8500-node.

### C. DER PFR under Generation Outage

This subsection studies the DER PFR responses to generation outage. The DER PFR is activated when the frequency deviates more than its PFR deadband (0.017 Hz in this study) after a generation loss. In the simulation, Gen 4 with 40-MW power output is tripped at the fifth second. Similar to the previous subsection, 20 DERs provide 10% of the AGC response, with each DER providing 0.5% AGC response. The rest of the AGC is provided by thermal units. Note that the loads are kept constant in this case for clear presentation. The DER intermittency varies near 0.8 MW.

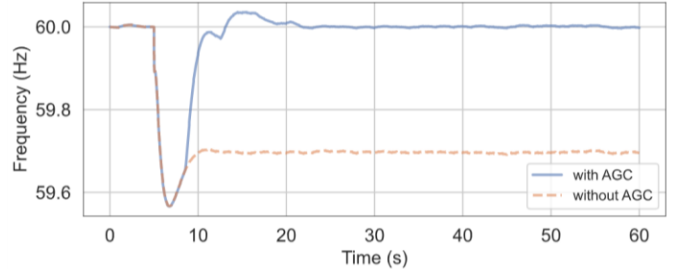
Fig. 12. Frequency response with/without AGC.

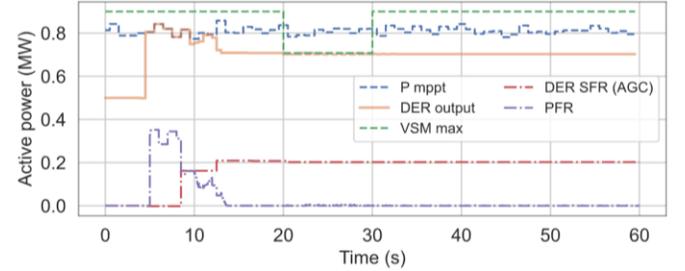
Fig. 13. DER power output, PFR, SFR, and its MPPT and VSM limits.

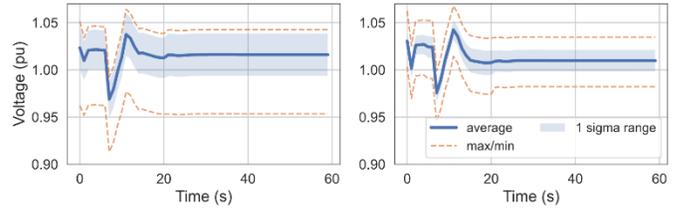
Fig. 14. Voltage response of (a) feeder 34-bus; (b) feeder 8500-node.

Fig. 12 compares the system frequency dynamic response with and without the AGC response. As expected, the frequency does not recover to 60 Hz without the SFR AGC, although it settles at a value less than 60 Hz. With the SFR AGC enabled, the frequency is restored to 60 Hz. Fig. 13 demonstrates the DER PFR and SFR after the generation outage, along with the DER actual power outputs and limits, including MPPT $P_{mppt}$ and VSM max. The total power output including the reference power (i.e., the DER's dispatched power output 0.5 MW), PFR, and SFR is less than its $P_{mppt}$ and VSM power limit; thus, these limits are respected by the DER's dynamic response. This figure also demonstrates that after the generation outage DER's PFR will response first to increase DER's power output to support the system frequency. Then DER's SFR starts to increase the power output to stabilize the system frequency. After the SFR supporting the frequency to a normal level, PFR will phase out. Fig. 14 shows two feeders' voltage behavior after the generation outage. At time 5 seconds, both feeders experience voltage dips, followed by small voltage overshoots that are mainly resulting from transmission voltage reacting to frequency regulations. The fact that all these behaviors are captured shows the effectiveness of the proposed framework.

### D. T&D Dynamic Co-Simulation Computational Performance

Previously, one major challenge of T&D dynamic co-simulation was the computational burden. The proposed T&D dynamic co-simulation model and framework leverage the efficient co-simulation platform HELICS, which parallels the distribution feeder QSTS simulation; therefore, the overall computational performance is high. In the case studies, for the IEEE 14-bus system with 34-bus and 8,500-node T&D networks, the 60-second dynamic simulation takes approximately 60 seconds on a personal laptop with Intel Core i7-10610-U processor. The case studies show that as the framework incorporates more detailed feeders, the computational time does not increase much accordingly. This demonstrates that the proposed co-simulation model is efficient and scalable to large T&D network dynamic simulations.

## V. CONCLUSIONS

DERs, including distributed PV, have been increasingly deployed in power systems. To better use PFR and SFR services provided by DERs, the dynamic response of DERs in T&D networks should be modeled. This paper proposed a T&D dynamic co-simulation framework to model the dynamic response of DERs providing PFR and SFR. The impacts of DER frequency responses on both the transmission system frequency response and the distribution feeder voltage are modeled. In addition, DER power variation regarding the maximum available power of distributed PV can be modeled as well; therefore, DER power uncertainty can be endogenously considered in the proposed dynamic co-simulation model.